# Use of Approaches to the Methodology of Factor Analysis of Information Risks for the Quantitative Assessment of Information Risks Based on the Formation of Cause-And-Effect Links

Ihor Dobrynin, Tamara Radivilova, Nadiia Maltseva, Dmytro Ageyev
Department of Infocommunication Engineering
Kharkiv National University of Radio Electronics
Kharkiv, Ukraine
ihor.dobrynin@nure.ua, tamara.radivilova@gmail.com, dmytro.aheiev@nure.ua

*Abstract*— The paper suggests methods to the assessment of information risks, which makes the transition from a qualitative assessment of information risks (according to the factor analysis of information risks methodology) to a quantitative assessment. The development factor analysis of information risks methodology of the methodology was carried out using the mathematical apparatus of probability theory, namely Bayesian networks. A comparative analysis of the standard factor analysis of information risks methodology and the developed methodology using statistical data was carried out. During the analysis, the cause and effect relationships of the confidentiality violation have been formed, defined and given in the corresponding table and in the form of the Ishikawa diagram. As an example, it was calculated the amount of risk the company may be exposed to in case of violation of information confidentiality according to the standard factor analysis of information risks methodology and the developed methodology. It is shown that the use of proposed technique allows quantifying the risk assessment that can be obtained using the factor analysis of information risks methodology.

## I. Introduction

In the modern world, information is transmitted and stored in infocommunication systems and is the most valuable product. The main component of infocommunication systems is the information protection system (IPS), from the correct functioning of which the security of information components depends. Method for ensuring the continuity of the IPS provides a stage of information security risk management [1]. Risk is an event that can cause loss or damage. Risk assessment is a process, as a result of which the damage or loss is determined in quantitative or qualitative terms [2].

In information security, risk assessment is a way of making decisions. The most common decision is the rationale for the costs of implementing a product to protect information. Protection is implemented, in order to avoid damage to valuable assets [3]. Prevention of damage occurs by neutralizing risks (through the implementation and management of measures and protective equipment). Risk assessment and management processes are the foundation for building the information security management system (ISMS) of organization. The effectiveness of these processes is determined by the accuracy and completeness of the analysis and risk assessment, as well as the effectiveness of the mechanisms used in the organization to make managerial decisions and control their implementation.

The processes of creating, constructing, implementation, operation, monitoring and improvement of ISMS presented both in international standards [3] (for example, the ISO/IEC 27000 standards line), as well as the development of well-known companies, for example: IT-Grundschutz, ISACA, etc.

However, the analysis of standards in the field of information security has shown that they do not clearly indicate the way in which these procedures should be carried out. That is, they do not give a concrete answer to what kind of methodology for risk assessment should be used. As a rule, this task is assigned to the heads of enterprises or persons responsible for the implementation and support of the ISMS. Thus, the task of selecting (developing) a methodology for assessing information risks, with its simplicity and visibility, the reliability of the results obtained with the help of it is an extremely important scientific and practical task [4, 5].

The aim of the work is to develop a methodology for quantifying information risks based on the methodology of factor analysis of information risks.

## II. Risk Assessment Methods

Today, there are a wide variety of methods for assessing information risks [6, 7]. The most commonly used methods are:

- quantitative methods (the risk is estimated through a numerical value, for example, the amount of expected annual losses);
- methods that use risk assessment at a qualitative level (for example, on a scale of "high", "medium", "low");
- methodologies using mixed estimates.

### A. Quantitative method

Quantitative risk assessment is used in situations where the threats and associated risks can be compared with the final quantitative values expressed in money, percentages, time, human resources, and so on. The method allows to obtain specific values of risk assessment objects when implementing threats to information security. In a quantitative approach, all elements of risk assessment are assigned specific and real quantitative values. The algorithm for obtaining these values should be clear and understandable. The object of valuation can be the value of the asset in monetary terms, the likelihood of the threat, the damage from the implementation of the threat, the cost of protective measures, and so on.

*B. Qualitative method*

Because of the large uncertainty, it is not always possible to obtain a specific expression of the object of evaluation. In this case, a qualitative method is used. Qualitative approach does not use quantitative or monetary expressions for the object of evaluation. Instead, the evaluation object is assigned an indicator ranked by a three-point (low, medium, high), five-point or ten-point scale (0 ... 10). To collect data with a qualitative risk assessment, questioning of target groups, interviewing, questioning, personal meetings are used. The analysis of information security risks by a qualitative method should be conducted with the involvement of employees with experience and competence in that area in which threats are addressed.

The aim of both approaches is to understand the real risks of the company's information security, to determine the list of current threats, and to select effective countermeasures and means of protection. [8, 9]. Each approach to risk assessment has advantages and disadvantages. The quantitative method provides a visual representation in the money for the objects of evaluation (damage, costs), however it is more laborious and in some cases is not applicable.

A qualitative method allows to perform a risk assessment faster, but the estimates and results are more subjective and do not provide a clear picture of the damage, costs and benefits of implementing ISMS [10]. It should be noted that at this time, the main approaches to information technology risk management are based on the requirements of the Information Management and Audit Standard Cobit v.5.0; Risk Management Guidelines for Information Technology NIST 800-30; ISO / IEC 27000 and ISO / IEC 31000 series standards [3], BSI-Standards from IT-Grundschutz, etc. The next methods are usually used directly for risk assessment: the method of assessing operational-critical threats; Operationally critical threats, assets and vulnerability evaluation (OCTAVE); Methodology for Risk Assessment by the National Institute of Standards and Technology (NIST); Risk Analysis and Management Method (CRAMM) and others. Recently increasing attention is paid to the method of factor analysis of information risks (FAIR), which provides the most complete consideration of the factors of information risks [4]. The analysis of known methods carried out by the authors showed that each method has both advantages and certain disadvantages.

From the companies management point of view the main disadvantages considered methods are the provision of a qualitative (but not quantitative) assessment, that does not give a specific risk value, with which the leaders of organizations could work (only approximate values are given, the range of which is quite variable and is indicated in the corresponding scales used) and taking into account the insufficient number of factors influencing the risk assessment.

Consider the main approaches of the FAIR methodology. This method is based on the analysis of factors that affect the various components of risk. According to this technique, first of all, the risk depends on frequency of occurrence of the incident and the probably losses of its occurrence and based on a cause-effect analysis. Cause-effect analysis is a combination of analysis of the "tree" of faults (causes) and the "tree" of events (effects), which is used to analyze complex problems that depend on a variety of reasons. A key aspect is the application of a cause-effect diagram, otherwise known as Ishikawa's Diagram.

The advantages of this approach are as follows: at the same time allows to analyze different categories of causes of the problem; encourages creativity when brainstorm; provides a visualized structured presentation of the causes of the complex problem; this method is used to analyze various ways of developing events that could take place in the system after a critical event, depending on the functioning of specific subsystems (for example, emergency systems). If these methods are quantified, they represent an estimate of the probability of the various possible consequences of a critical event.

### III. PROPOSED METHOD

We will carry out a transition from a qualitative assessment of information risks (by the method of FAIR) to quantitative assessment. At the same time, we will take into account the requirements of the international standard ISO / IEC 27001: 2013, as the most relevant in the field of information security. This transition involves selection of an asset for which a risk assessment will be carried out. Note that as an asset, anyone for which the risk assessment is carried out can be selected. In the work as an asset, the authors considered a file containing confidential information (restricted access information) and located on the computer.

In accordance with the standard ISO/IEC 27001:2013, the violation of information security (IS) is a violation of confidentiality (C), integrity (I) and availability (A) of information, at this stage determine:

- possible events leading to a violation of C, I, A. Note that the definition of these events must be carried out separately for each of the IS properties, that is, separately for C, I, A;
- determines at what expense this event may occur;
- determines the reasons that may lead to these events.

The results of the second stage for the formation and determination of cause-effect relationships should be placed in the appropriate tables or in the form of Ishikawa graphs for each of the properties IS: C, I, A.

It should be noted that when forming the corresponding tables, it is necessary to consider both the elements of the FAIR method and the requirements of the standard ISO/IEC 27001:2013 [3], that are: possible event and its probability is equal to contact; conditional probability of occurrence of an event is equal to action; causes that give rise to hypotheses and the relevant paragraph of the standard ISO/IEC 27001:2013 is equal to level control; through which the event can happen (hypothesis) and its probability is equal to possibilities threat. Together with the reasons, which lead to the emergence of hypotheses, it is advisable to refer to relevant paragraphs of ISO/IEC 27001:2013 [3]. These are elements and mechanisms that need to be implemented in the organization to effectively ensure the information security.

The next step provides for the definition of probabilities for implementing hypotheses and conditional probabilities of events occurrence (columns 2 and 4 of Table 1, respectively). These probabilities are obtained based on a priori known statistical data of analytical companies that are professional in this field and relevant decisions of the company's experts. Thus, the data shown in the table allow us to calculate the probability of realizing threats through specific vulnerabilities.

Consider the events that can lead to violation of confidentiality, integrity or accessibility of information. Obviously, they are independent, that is, the appearance of one of these events does not affect the appearance of another. In the presented table, these events are listed in the "Possible event and its probability" column. The implementation of any of these events (factors) leads to loss of confidentiality, integrity or availability of information with a probability $P$.

Based on the theorem, that the probability of occurrence of at least one of the events, independent in the aggregate, is equal to the difference between unity and product of probabilities opposite events [11], we obtain the expression $P = 1 - \prod_{i=1}^{n}(1 - P(A_i))$, where $n$ is the number of events that may violate confidentiality, integrity or availability; $i$ is a current event number; $P(A_i)$ is probability of realization of the event.

It is advisable to take into account the fact that the realization of the events of $A_i$ also depends on a number of factors (or hypotheses). Let's mark their probabilities of implementation through $P(H_{ij})$. They are also independent of each other in the same event. Denote by $P(A_i | H_{ij})$ the conditional probability of occurrence of the event $A_i$ under the condition of the $j$-th hypothesis. Then, based on the formula of the complete probability and the formula of adding probabilities [11], we obtain the expression $P(A_i) = \sum_{i=1}^{n}\sum_{j=1}^{m} P(H_{ij})P(A_i | H_{ij})$, where $i$ is the current event number, the $j$ is current hypothesis number, the $n$ is the number of appropriate events, $m$ is a number of hypotheses. Thus, the application of the above formulas will give the value of the probabilities of confidentiality violations, integrity and availability separately. The expected value of the loss amount can be calculated by the formula: $R_i = \sum_{i=1}^{n} P_i \cdot E_i$, where $R_i$ is probability of violation of C, I, A; $E_i$ is the size of losses from the onset of these events.

The Ishikawa diagram allows to identify and systematize various factors and conditions that affect the assessment of the risk of confidentiality violation (Fig.1). The table representation of active (file located on local host) is presented in Table 1.

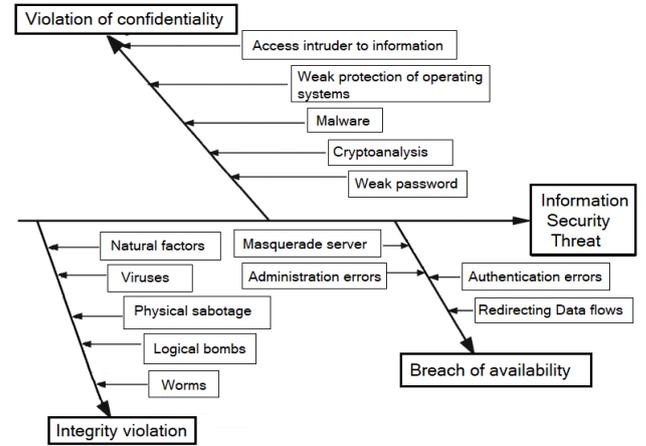

Fig. 1. Possible causes on Ishikawa diagram

## IV. ANALYSIS OF PROPOSED METHOD

As part of the study, the authors conducted a comparative analysis of the standard FAIR risk assessment methodology and the developed approach using the statistics provided by leading companies to assess the threats to information security. As an asset, a file with confidential information was considered. The actions that could damage the company were considered by the attacker to violate the privacy, integrity and availability of the information stored in this file. According to the FAIR methodology, the risk is the product of the probabilistic frequency of insured events and the probable value of possible losses. As noted above, the risk on the asset is the products sum of the probability values of violation of confidentiality, integrity and accessibility of information on the amount of possible damage from the onset of these events. As an example, the magnitude of risk was calculated, which the company may experience in case of violation of the confidentiality of information. After establishing all the necessary data $P(A_i)$, we calculated total probability of violation of the confidentiality $P$, the losses that can be incurred by the company in violation of the confidentiality of information Ekonf. To do this, we will estimate the amount of loss in case of violation of the confidentiality of information for each form of loss. As noted it possible to take into account the loss as the sum of losses for all types of damage. Risk of confidentiality $R$ calculate according to the formula (1). Thus, in accordance with the proposed methodology, the quantitative risk value for an asset was obtained in case of violation of its confidentiality. According to the standard method, we can say that the risk is average, based on the low frequency of events that lead to losses (between 1-0.1 times a year) and the average probable value of losses (between 1000 and 10000 USD).

TABLE I. SOME EVENTS THAT MAY BE CAUSE A VIOLATION OF CONFIDENTIALITY OF INFORMATION

| Possible event and its probability $P(A1_i)$ | Due to what the event (hypothesis) and its probability may occur, $P(H1_{ij})$ | The reasons for the hypothesis and the corresponding paragraph of the standard ISO/IEC 27001:2013 | Conditional probability of occurrence of an event, $P(A1_i|H1_{ij})$ |
|---|---|---|---|
| Violation of the confidentiality of information due to imperfection of physical protection $P(A1_1)$ | Overcoming an intruder perimeter of the enterprise security zone $P(H111)$ | Insufficient effectiveness of measures aimed at ensuring the physical protection of the perimeter, building, windows and doors, (A.11.1) | Violation of the confidentiality of information when overcoming the perimeter of the security zone of the enterprise, $P(A11|H111)$ |
| | Failure of an enterprise policy "clean table", $P(H112)$ | Failure to comply with clean desktop and screen policies, negligent attitude to the preservation of documents with restricted access, (A.11.2.9) | Violation of confidentiality at the failure of an enterprise policy "clean table", $P(A11|H112)$ |
| Violation of confidentiality by not effectively control access $P(A1_2)$ | Getting an attacker to access restricted information due to incorrectly configured host privileges, $P(H121)$ | There is not enough effective policy to delimit the access to the local host, (A.9.2, A.9.4) | Violation of the confidentiality of information when an attacker receives access to restricted information due to improperly configured host privileges. $P(A12|H121)$ |
| | Realization of password hacking, $P(H122)$ | Insufficiently effective password management system (easy guessing passwords, insufficient frequent replacement), (A.9.4.3) | Violation of confidentiality at hacking password protection, $P(A12|H122)$ |
| | Removing the restricted information from removable storage, $P(H123)$ | Insufficient effectiveness of measures aimed at ensuring the safe operation of removable storage, (A.8.3) | Violation of confidentiality while removing the restricted information from removable storage, $P(A12|H123)$ |
| | Bribery of employees by an attacker, $P(H124)$ | Bribery of employees, (A.7) | Violation of confidentiality at bribing employees of the organization, $P(A12|H124)$ |
| | Unintentional harm of employees, $P(H125)$ | Unintentional destructive actions of employees, (A.7) | Violation of confidentiality at the unintended harm the organization's staff, $P(A12|H125)$ |

Thus, the risk value obtained with the proposed method falls within the range of the probable loss value, which is determined by the standard approach. However, unlike a qualitative assessment of the FAIR technique, the resulting risk value is quantitative, which allows more efficiently predict the costs of information security.

It should be noted that when calculating the developed methodology it is necessary to respond responsibly to the process of threats identification and the reasons for their occurrence (see Table 1), and take into account company and analytical agencies statistics. If necessary, use the experts services.

V. CONCLUSION

The proposed approach to the assessment of information risks is based on the concept of the FAIR technique, taking into account the requirements of the international standard ISO/IEC 27001:2013, as the most current in the field of information security, and allows quantifying the risk assessment that can be obtained through the qualitative method of FAIR. The methodology was developed using the mathematical apparatus of probability theory, namely Baesian networks. To determine the risk, a causal-effect analysis was used.

The results of the work should be used to quantify the information risks in various companies and organizations that create or operate the SMB. Further development of work should be research aimed at increasing the level of expert information used in the stage of determining the likelihood of the implementation of hypotheses and conditional probabilities of occurrence of events.